\documentstyle[aps,prb]{revtex}
\input epsf.sty
\begin{document}
\twocolumn[\hsize\textwidth\columnwidth\hsize\csname@twocolumnfalse\endcsname

\draft

\title{Effective mass staircase and the Fermi liquid parameters\\ 
for the fractional quantum Hall composite fermions}
\author{Masaru Onoda$^{1}$, Takahiro Mizusaki$^{2}$, 
and Hideo Aoki$^{3}$}
\address{$^{1}$Department of Applied Physics, University of Tokyo,
Bunkyo-ku, Tokyo 113-8656, Japan}
\address{$^{2}$Department of Law, Senshu University, 
Tama, Kawasaki, 214-8580, Japan}
\address{$^{3}$Department of Physics, University of Tokyo,
Hongo, Tokyo 113-0033, Japan}
\date{\today}

\maketitle

\begin{abstract}
Effective mass of the composite fermion in the fractional quantum Hall 
system, which is of purely interaction originated, is shown, 
from a numerical study, to exhibit a curious nonmonotonic behavior 
with a {\it staircase} correlated with the number ($=2,4,\cdots$) of 
attached flux quanta. This is surprising since 
the usual composite-fermion picture predicts a smooth behavior.  
On top of that, significant interactions are shown to 
exist between composite fermions, where the 
excitation spectrum is accurately reproduced in terms of 
Landau's Fermi liquid picture with negative 
(i.e., Hund's type) orbital and spin exchange interactions.
\end{abstract}

\pacs{73.43.-f}

]

{\it Motivation.---} 
Although the composite-fermion (CF) picture\cite{CFs} 
for the fractional quantum Hall (FQH) system provides a 
fascinating way of understanding the 
electron system interacting in a two-dimensional space 
in a strong magnetic field, there still remain fundamental 
properties of the CF that need to be revealed.  
Indeed, the mass, the most important quantity to 
characterize a particle, is far from fully understood 
for the composite fermion.  There is a good reason for this --- 
unlike usual many-body systems where we can talk about effects 
of the interaction on the kinetic energy, 
the FQH system has a singular and intriguing situation 
where the kinetic energy is quenched (with the bare mass being ``infinite") 
so that a finite effective mass is purely interaction-originated when the 
magnetic field is strong enough and a single Landau level may be considered.  

Hence the effective mass has to be put in by hand in the composite fermion 
picture.  In this picture the incompressible FQH liquid 
for odd-fraction Landau-level filling $\nu=p/(\tilde{\phi}p\pm 1)$ 
($p$: an integer) is regarded, in the mean-field sense, 
as an integer quantum Hall state of CF's 
with $\nu^* = p$, where a CF is 
an electron attached with an even number, $\tilde{\phi}$, of 
filamentary flux quanta.  Since the external magnetic field is partly 
sucked by electrons, each CF feels, on average, a reduced magnetic field, 
$eB^* = eB - 2\pi \tilde{\phi}n_e$ 
where $n_e$ is the number density of electrons, 
and the Landau level spacing becomes an effective cyclotron energy 
$\omega_c^*=eB^*/m_0^*$ (in natural units with $\hbar=c=1$) 
with $m^*_{\rm 0}$ being the effective mass in question.   
However, the system of CF's reduces to the original one only 
in the mean-field sense, and fluctuations around the mean 
field, being a gauge field \cite{PRB44_05246_91}, should be strong 
due to the absence of small parameters such as interaction/kinetic energy.  
This should be especially so at even fractions, 
where the excitation gap $\propto B^*=0$ vanishes.  
In a seminal paper\cite{PRB47_07312_93}, 
Halperin, Lee, and Read systematically developed the CF theory. 
However, the effective mass remains to be difficult 
quantity to estimate, except for a dimensional argument 
that $eB^*/m^* \sim e^2/\ell$ where $\ell$ is the magnetic length.   

Another fundamental question is: the CF picture, in its naive 
form, does not say anything about the electron-electron 
repulsion, which is the very origin of the FQH states.  
This flaw has recently been remedied by considering 
a ``dipole" CF, where the flux attachment is regarded as mimicking 
the correlation hole due to Coulomb repulsion between 
electrons \cite{PRL62_00086_89}. 
Still, there should be residual interactions between CF's, 
which may be strong, since the strong gauge field fluctuations 
should be reflected on the interaction.   Stern and Halperin have constructed 
the Fermi liquid theory of CF's with a singular Landau function 
to renormalize the CF mass consistently\cite{PRB52_05890_95}.
However, the validity of the perturbative treatments of 
the residual interaction between CF's as well as 
the quantitative estimate of the CF effective mass
are still some way from a full understanding 
due to the quenched kinetic energy. 
Hence numerical studies are valuable as an approach 
complementary to analytic ones\cite{PRL72_00900_94,Sitko}.

In our recent work\cite{PRL84_03942_00}, 
we have questioned, as a crucial test for a many-body theory, 
whether the CF picture can reproduce low-lying excitation spectra.  
We have shown, from an exact-diagonalization for finite systems, 
that there exists a striking 
one-to-one correspondence in the low-lying excitation spectra 
between the exact result and the free CF system in zero magnetic field 
at even-fraction $\nu$'s.   This has enabled us to
estimate the CF effective mass from the 
lowest excitation energy, by assuming free CF's, to show that $m^*$ becomes 
heavier as we go to higher even fractions 
($\nu=1/2\rightarrow 1/4\rightarrow 1/6$).  

However, we definitely need to examine this more systematically --- 
(i) First, we want to study $m^*$ for general fractions, including odd 
fractions on which Laughlin's quantum liquid resides.  
(ii) Second, the residual CF interaction should be probed, in terms of 
Landau's Fermi liquid parameters, 
especially in the presence of spin degrees of freedom.  
These are exactly what we have done here 
with the numerical (Lanczos) diagonalization for finite systems 
in the subspace projected to the lowest Landau level.  We shall show that 
(i) the CF effective mass exhibits a curious 
{\it staircase} against $\nu$, which is 
surprising since the usual CF theory would predict 
a smooth dependence.  
(ii) Low-lying excitation spectra indicate that 
there are indeed significant interactions between CF's, 
which can be accurately described, in finite systems, 
in terms of Landau's Fermi liquid parameters with {\it negative} 
(i.e., Hund's type) spin {\it and} orbit exchange interactions.

For the numerical study the spherical system is adopted, following Haldane 
\cite{PRL51_00605_83}.  
Dirac's quantization condition dictates that
the total flux $4\pi R^2 B$ be an integral ($N_{\phi}$) multiple of 
the flux quantum, where $R$ is the radius of the sphere. 
The eigenvalue of the noninteracting Hamiltonian is
$
\varepsilon 
= 1/(2mR^2)[l(l+1)-\left(N_{\phi}/2\right)^2],
$
where $l\ge N_{\phi}/2$ is an integer with the equality 
corresponding to the lowest Landau level (LLL).
The electron-electron interaction is the 
whole Hamiltonian in the LLL subspace.  
The Zeeman effect is neglected for simplicity,
so the system has the SU(2) symmetry.

{\it Free composite fermion analysis. ---} 
For the FQH states at odd fractions, 
$
\nu = 2\pi n_e/(eB) = d_s p/(d_s\tilde{\phi}p\pm 1),
$
the effective mass can be estimated from the the excitation gap, 
i.e., $\omega_c^*$ in the CF system 
with 
$
\nu^*\equiv 2\pi n_e/(eB^*)=d_s p.
$
Here $p$ is a positive integer and
$d_s$ is the spin degeneracy ($d_s=1$ for the spinless case, 
$d_s=2$ for the spinful case). 
For the spinful case we take the simplest definition, 
i.e., attaching the same number ($\tilde{\phi}$: even) of flux quanta 
to up and down spin electrons.\cite{PRB51_04347_95} 

For finite systems we have to specify the number of 
fluxes for a given number of electrons, $N_e$, and $\nu$ with some care.  
This can be done in terms of the CF picture.  
The number of flux $N^*_{\phi}$ left behind after 
the flux attachment is 
$
N^*_{\phi}=|N_{\phi}-\tilde{\phi}(N_e-1)|.
$
For the integer QH state of CF's, $N$ and $N^*_{\phi}$ must satisfy 
$
N_e= d_s p(N^*_{\phi}+p),
$
because CF's fill up to the $p$-th effective Landau level
for each spin.  Then $N_{\phi}$ is given by
$
N_{\phi} = (d_s\tilde{\phi}p\pm 1)N_e/(d_s p)-(\tilde{\phi}\pm p),
$
where $\pm$ corresponds to $\nu^{<}_{>}1/\tilde{\phi}$.  
The energy gap in the mean field is $\Delta \approx \omega_c^*$, 
but on a finite sphere the precise expression is
\begin{equation}
\Delta 
=\frac1{m^*_0R^2} \left(\frac{N^*_{\phi}}2 + p \right)
=\frac{4\pi n_e}{m^*_0}\cdot \frac{N_e+d_s p^2}{2d_s p (N_e-1)}.
\label{OFgap}
\end{equation}
Note that, with $R$ scaling as $4\pi n_e R^2 = N_e-1$, 
$B$ (hence $B^*$) are functions of $N_e$ to 
satisfy Dirac's quantization.
We can obtain a similar expression for 
the metallic case with $\nu=1/\tilde{\phi}=$even fractions 
with $B^*=0$ by putting $N^*_{\phi}=0$ ($N_e=d_s p^2$). 
This time we can look at how the gap vanishes for $N_e\rightarrow \infty$, 
which has a scaling,\cite{PRL84_03942_00} 
\begin{equation}
\Delta 
=\frac{l_F+1}{m^*_0 R^2}
=\frac{4\pi n_e}{m^*_0} \cdot\frac{\sqrt{N_e}}{\sqrt{d_s}(N_e-1)},
\label{EFgap}
\end{equation}
as a one-exciton (one electron-hole pair) excitation 
if we concentrate on the closed-shell case\cite{PRL84_03942_00} 
with $N_e/d_s=(l_{\rm F}+1)^2$ where $l_{\rm F}$ 
is the Fermi angular momentum. 

{\it Effective mass against $\nu$. ---} 
Let us first estimate the CF effective mass in the noninteracting 
CF picture, eqs.(\ref{OFgap}) and (\ref{EFgap}), 
for various $N_e$ for each value of Landau-level filling, 
in both the spinless and spinful cases.  
The maximum system size has
the dimension of the Hamiltonian matrix
as large as 23 million.
We have varied the magnetic field for each value of the number of electrons 
to have various $\nu$, so that we have normalized the mass 
by $e^2/(4\pi n_e)^{1/2}$, 
since this way the plot of $m_0^*$ versus $\nu$ may be regarded as 
representing the magnetic field dependence of $m_0^*$.

\begin{figure}[h]
  \begin{picture}(200,100)
    \put(0,0){\epsfxsize 160pt \epsfbox{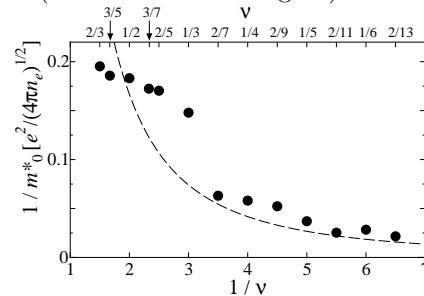}}
  \end{picture}
  \caption{Inverse effective mass estimated 
        from the size scaling of the excitation energy 
	in the spinless system ({\large $\bullet$}). 
	The dashed line is the CF mean-field result. 
  }
\label{spinless_mass}
\end{figure}

In Fig.\ref{spinless_mass} which plots the 
inverse effective mass against $\nu$ for the spinless case, 
we can immediately see that the effective mass is very nonmonotonic 
with a step-function-like behavior 
each time the Landau-level filling 
passes the boundary across difference numbers of attached flux 
quanta, i.e., $\tilde{\phi} \to \tilde{\phi}+2$.  
This is totally unexpected, since the CF theory, with 
its mean-field treatment after a certain canonical transformation, 
predicts a smooth function 
(dashed line in the figure)\cite{MSmass}.  

Within each step, $m^*_0$ is only weakly dependent on $\nu$, 
which implies that the effective mass 
is basically determined by $\tilde{\phi}$, the number of attached fluxes. 
Note that the usual practice of linearly plotting $\omega_c^*$ 
versus $B^*$ is allowed only when this property exists.  
Experimentally, Leadley {\it et al.} estimated the effective mass 
from the thermal activation energy in the Shubnikov-de Haas oscillation 
and observed difference $B$-dependence of sequences 
for $\tilde{\phi}=2,4$\cite{PRL72_01906_94}.  
Curiously, if we look at the positions of the steps, 
the result dictates that we have to 
attach exactly two flux quanta ($\tilde{\phi}=2$) at $\nu=1/3$, while 
we could have chosen from two or four since $\nu=1/3$ is 
just the boundary between the two choices in the CF picture.  
We are referring to a large gap in $1/m^*_0$ between $\nu=1/3$ 
and $2/7$, but, to be precise, whether 
this is a jump or a continuous curve will have to be studied.  
For the sample size studied here, we see only a shoulder
around $\nu \sim 1/5$ because the energy scale becomes
small there.

\begin{figure}[h]
  \begin{picture}(200,100)
    \put(0,0){\epsfxsize 160pt \epsfbox{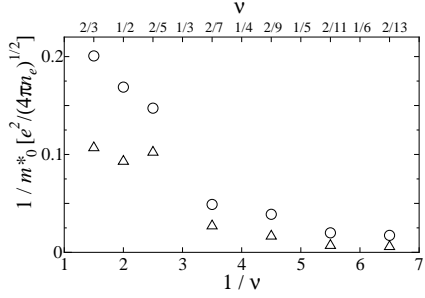}}
  \end{picture}
  \caption{Inverse effective mass estimated 
        from the system size scaling of the 
        spinflip ({\scriptsize $\triangle$}) or 
	no-spinflip ({\large $\circ$}) excitation energies 
	in the spinful system. 
  }
\label{spinful_mass}
\end{figure}

{\it Spin effect. ---} 
Figure \ref{spinful_mass} plots 
the $1/m^*_0$ when the spin degrees of freedom are included.  
We can see that 
$m^*_0$ appears to be about twice as heavy as that of the spinless system 
in Fig. \ref{spinful_mass} 
when we estimate the mass from spin-flip ($\Delta S\neq 0$) 
excitations (which happen to be the lowest one), while 
the effective mass estimated from 
excitations that involve no spin flips 
is close to that for the spinless result above.  
This indicates that we should not stick to the free CF picture, 
but that spin-dependent (exchange) interactions between CF's have to 
be considered.

{\it Fermi liquid description. ---} 
So we turn to the Fermi liquid picture of CF's 
for even-fraction metals. 
This picture assumes that the excitation energy, $\delta E$, is given 
as a functional of the deviation in the particle number (in a finite system) 
from the ground state, 
$\delta N_{\mbox{\boldmath{$l$}}\sigma}$ and 
the Landau function, 
$f_{\mbox{\boldmath{$l$}}\sigma\:\mbox{\boldmath{$l$}}'\sigma'}$ 
which is labeled for a sphere by the angular momentum 
$\mbox{\boldmath{$l$}}$ and the spin.   

If we expand the Landau function to the first order in
$\mbox{\boldmath{$l$}}\cdot\mbox{\boldmath{$l$}}'$ 
(which corresponds to the spherical harmonics expansion for 
$f_{\mbox{\boldmath{$p$}}\:\mbox{\boldmath{$p$}}'}$ 
in a flat system)  and
$\mbox{\boldmath{$\sigma$}}\cdot\mbox{\boldmath{$\sigma$}}'$, 
we have
\begin{eqnarray}
\delta E 
&=& \sum_{\mbox{\boldmath{$l$}}\sigma}
\tilde{\xi}_{\mbox{\boldmath{$l$}}}\delta N_{\mbox{\boldmath{$l$}}\sigma}
+\frac12\sum_{\mbox{\boldmath{$l$}}\sigma}\sum_{\mbox{\boldmath{$l$}}'\sigma'}
 \biggl[(f_0+g_0\mbox{\boldmath{$\sigma$}}\cdot\mbox{\boldmath{$\sigma$}}')
\nonumber\\
&&
+(f_1+g_1\mbox{\boldmath{$\sigma$}}\cdot\mbox{\boldmath{$\sigma$}}')
   \frac{\mbox{\boldmath{$l$}}\cdot\mbox{\boldmath{$l$}}'}
   {(l_{\rm F}+1)^2}
   \biggr]
 \delta N_{\mbox{\boldmath{$l$}}\sigma}
\delta N_{\mbox{\boldmath{$l$}}'\sigma'},
\label{Landau_fn}
\end{eqnarray}
where 
$\tilde{\xi}_{\mbox{\boldmath{$l$}}}
= \mbox{\boldmath{$l$}}^2/(2m^*_{\rm FL}R^2)-\mu$
with $\mu$ being the chemical potential.
In order to pick up the scale independent variables, 
it is convenient to introduce dimensionless parameters,
$
F_i=2d_s R^2 m^*_{\rm FL}f_i, \; G_i=2d_s R^2 m^*_{\rm FL}g_i,
$
where $m^*_{\rm FL}$ is the effective mass introduced in the context of
the Fermi liquid theory.  
The definition reduces, for $N_e\gg1$, to the usual one in 2D, 
$F_i = 4\pi R^2 D_{\rm F} f_i$ and $G_i = 4\pi R^2 D_{\rm F} g_i$,
where $D_{\rm F}=d_s m^*_{\rm FL}/(2\pi)$ is 
the density of states per unit area.
Since the Fermi liquid theory is meant to be applied to 
low-lying excitations, we consider excitations near the Fermi energy 
i.e., $\delta N_{\mbox{\boldmath{$l$}}\sigma}$ with $l\sim l_F$.
The lowest-lying excitations 
for the closed-shell case is the one particle-hole pair with 
$(\delta N_{\mbox{\boldmath{$l$}}_1\sigma_1},
\delta N_{\mbox{\boldmath{$l$}}_2\sigma_2}) = (1, -1), 
l_1=l_{\rm F}+1, l_2=l_{\rm F}$.  
So we have
\begin{eqnarray}
\delta E 
&=& 
\Delta_{\rm FL} 
\biggl[1+\frac1{d_s(l_F+1)}\Bigl\{(G_0+G_1)
\mbox{\boldmath{$S$}}\cdot\mbox{\boldmath{$S$}}
\nonumber\\
&&
+\frac14[F_1+G_1(3-2\mbox{\boldmath{$S$}}\cdot\mbox{\boldmath{$S$}})]
 \frac{\mbox{\boldmath{$L$}}\cdot\mbox{\boldmath{$L$}}}
{(l_{\rm F}+1)^2}\Bigr\}\biggr],
\end{eqnarray}
where $\Delta_{\rm FL}=(l_{\rm F}+1)/(m^*_{\rm FL}R^2)$,
$\mbox{\boldmath{$L$}}=\mbox{\boldmath{$l$}}_1-\mbox{\boldmath{$l$}}_2$
with $1\le L \le 2l_{\rm F}+1$\cite{commentL1}, and
$\mbox{\boldmath{$S$}}=\frac{1}{2}
(\mbox{\boldmath{$\sigma$}}_1-\mbox{\boldmath{$\sigma$}}_2)$.
Since the ground state is rotationally invariant for the closed-shell, 
$\mbox{\boldmath{$L$}}$ coincides with 
the total angular momentum of the excited state.
For the spinful case, the ground state has zero total spin 
and $\mbox{\boldmath{$S$}}$ is the total spin of the excited state.

So we can estimate the effective mass $m^*_{\rm FL}$
and the Landau parameters
by applying the above expression to the numerical results.
We can immediately make a qualitative observation that the Landau parameter
$G_0+G_1$ (the coefficient of
$\mbox{\boldmath{$S$}}\cdot\mbox{\boldmath{$S$}}$)
is {\it negative}.  This is readily seen from Fig.\ref{spinful_mass}, 
where $1/m^* \propto \delta E$ for spin-flip excitations is 
significantly smaller (about one half, which implies that $G_0+G_1 \sim -1$ 
for $N_e=8$).  For $\nu = 2/3$, $2/5$, $2/7$ (even numerators), 
the system is not a metal but an integer QH state of CF's (for which the 
ground states turn out to be spin-singlet).  So we 
are talking about $\omega_c^*$ in these cases, and 
the apparent reduction in $1/m^*_0$ in Fig.\ref{spinful_mass} again 
indicates a negative spin exchange correction to $\hbar \omega_c^*$. 

These observations are consistent with our 
previous  numerical result\cite{matsue}, 
which shows that (a) the spin-resolved excitation spectra resemble 
those for the free system with a shift 
$\propto -\mbox{\boldmath{$S$}}\cdot\mbox{\boldmath{$S$}}$, and 
(b) the ground-state total spin when we go away from 
the closed-shell case is correctly explained by the Hund's coupling.  
So we can conclude that the interaction between spinful CF's
has an exchange interaction of Hund's type
\cite{PRL73_03568_94}.

It is noted that there have been
some numerical indications such as those obtained
by Rezayi and Read \cite{PRL72_00900_94} in which Hund's rule is reported 
for the orbital angular momenta.
However, existence or otherwise of Hund's rule is a subtle
problem, especially for long-range repulsive interactions,
and the situation should be far from well-understood,
especially as regards the interaction between CF's with different spins.

\begin{figure}[h]
  \begin{picture}(200,170)
    \put(0,0){\epsfxsize 180pt \epsfbox{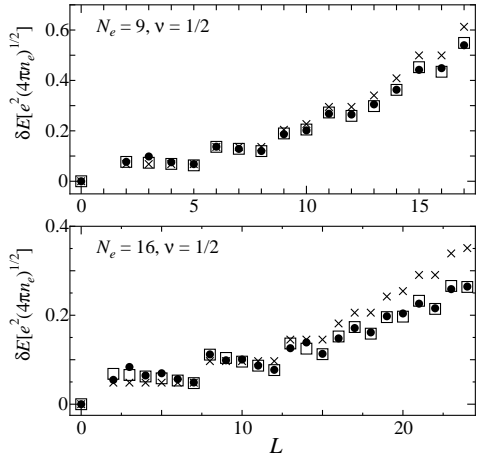}}
  \end{picture}
  \caption{Low-lying excitation spectra of 
        the $\nu=1/2$ system with $N_e=9$, $16$ 
        in the exact diagonalization ({\large $\bullet$}),
        the Fermi liquid theory ({\normalsize $\Box$})
	and the free CF model ($\times$).
	The best-fit effective mass and the Fermi liquid 
        parameter are 
	$[\sqrt{4\pi n_e}/(e^2m^*_{\rm FL}),
	F_1^{\rm spinless}] = [0.21,-0.8]$ for $N_e=9$, 
	$[0.27, -1.5]$ for $N_e=16$. 
         }
\label{spectra}
\end{figure}

{\it A quantitative estimate. ---} 
Can we quantify the Fermi parameters? 
If we go back to the spinless case for simplicity, 
the excitation energy is 
\begin{equation}
\delta E = 
\sum_{\mbox{\boldmath{$l$}}}
\tilde{\xi}_{\mbox{\boldmath{$l$}}}
\delta N_{\mbox{\boldmath{$l$}}}
+\frac{F_1^{\rm spinless}}{4m^*_{\rm FL}R^2}\cdot
\frac{\mbox{\boldmath{$L$}}\cdot\mbox{\boldmath{$L$}}}{(l_F+1)^2}
\label{spinless}
\end{equation}
with $F_1^{\rm spinless} = F_1 + 3G_1$, from which 
$m^*_{\rm FL}$ and $F_1^{\rm spinless}$ can be estimated.  
So here we have attempted to best fit 
$F_1$ and $m^*_{\rm FL}$ simultaneously with the least-square method.  
Figure \ref{spectra} is the fit ($\Box$) for 
the exact diagonalization result ({\large $\bullet$}) of
the low-lying ($-1<l-l_F<2$) excitation spectrum in the $\nu=1/2$ system
with each value of sample size, $N_e=9, 16$.
For each size the fit, with only two free parameters, is remarkably good 
for both the overall and shell\cite{PRL84_03942_00} structures.  
In this sense the system may be regarded as a liquid of CF's 
as far as the finite systems considered here are concerned. 
The fit is also good for $\nu=1/4$ (not shown here).  
In each case, $m^*_{\rm FL}$ is comparable with
$1/m^*_0$ estimated from the free CF picture.  
For example, in the $\nu=1/2$ case,
$\sqrt{4\pi n_e})/(e^2m^*_{\rm FL}) \simeq 0.21\:(0.27)$ 
for $N_e=9\:(16)$ against $0.18$ for $1/m^*_0$.

As for the Landau parameter $F_1^{\rm spinless}$, we can immediately 
tell that $F_1^{\rm spinless}$ is {\it negative}, 
if we compare the exact result ({\large $\bullet$}) with 
the free (i.e., $F_1^{\rm spinless}=0$) CF result ($\times$): 
the exact result lies significantly below the 
free CF result for larger $L$, which implies 
that CF's have an orbital exchange coupling of Hund's type.  
This observation is consistent with the Hund's rule
for the $\nu=1/{\rm even}$ ground state\cite{PRL72_00900_94}.
So the FQH system has spin- and 
orbital-exchange couplings both of which are Hund's type.   

Quantitatively, however, $F_1^{\rm spinless}$ is ill-behaved in the fitting 
if we assume the quantity is size-independent.  This is precisely 
why we had to fit $F_1^{\rm spinless}$ and $m^*_{\rm FL}$ 
for each value of $N_e$, and we end up with 
$F_1^{\rm spinless}=-0.8\:(-1.5)$ for $N_e=9\:(16)$.  
This leads us to question the assumption 
that the Landau parameters in the present system are scale invariant.  
The anomalous behavior should precisely be related to the 
infinite bare mass singularity of the FQH system.  
When the system had a finite bare mass $m$,
we would have a relation,
$
m^*_{\rm FL}/m = 1+F_1^{\rm spinless}/2
$
(with $1/2$ due to the two dimensionality).  
Since the LLL projected model has no bare kinetic energy to start with, 
the bare mass in the above relation, if at all meaningful, 
should diverge as $N_e \to \infty$ \cite{divmass}.  
Then $F_1^{\rm spinless}$ has to tend to 
$-2$, an anomalously large value. 
Then, how can we understand these results? 
This is exactly what we have addressed here,
and the conclusions are summarized as follows.
(a) The effective mass estimated
by comparing the exact result with that of a free system
is numerically close to those estimated with the Fermi liquid approach
and has a good property in the system size scaling.
The topology of the shell structure is also reproduced 
with the free CF picture.
In this sense the mean field picture seems to be all right.
(b) While the Fermi liquid picture with best-fit Landau parameters
does reproduce quantitative features of the excitation
spectrum, the system deviates from an `ordinary' Fermi liquid in
that Landau parameters are system-size dependent.

The best-fit $F_1^{\rm spinless} = -0.8\:(-1.5)$ for $N_e=9\:(16)$ 
does have a significant size dependence, although 
the system sizes considered here are too small to allow a scaling 
argument. In the thermodynamic limit, the 
Landau function could possibly be singular, 
as is considered by Stern and Halperin 
by summing the diagrams as required from the Ward-Takahashi identity.  
If the size-dependence of $F_1^{\rm spinless}$ 
has a logarithmic asymptote in the thermodynamic limit, then 
this may be related to 
the marginal Fermi liquid predicted by Ref.\cite{PRB47_07312_93}.
If the Fermi or marginal Fermi liquid persists in the thermodynamic limit, 
this would serve as an instance in which a system that 
has no small parameters 
(interaction/kinetic energy$=\infty$, $\tilde{\phi} \sim O(1)$) 
can be a Fermi liquid. 
Another interesting problem is the 
effect of the Landau level mixing.  These will serve as future problems.



\begin{references}
\bibitem{commentL1} 
For the reason why the one-exciton state with $L=1$ is absent 
in the numerical result, see \protect\cite{PRL84_03942_00}
\bibitem{divmass} 
This is not to be confused with the logarithmically diverging mass 
in Ref.\protect\cite{PRB47_07312_93}, where the effective mass is defined for 
the one-particle Green's function, not for 
excitations (a two-particle property).
\end{references}
\end{document}